\documentclass[11pt,twoside]{article}
\usepackage{amsmath}
\usepackage{amssymb}
\usepackage{hyperref}

\newcommand{\funcs}{\F^\infty_1}
\newcommand{\LPerf}{{\tt Perf}}
\newcommand{\Sel}{{\tt Sel}}
\newcommand{\Neigh}{{\tt Neigh}}
\newcommand{\Bene}{{\tt Bene}}
\newcommand{\Neut}{{\tt Neut}}
\newcommand{\SelNB}{{\tt SelNB}}

\newcommand{\SQD}{\mbox{SQD}}
\newcommand{\SLC}{\mbox{SLC}}
\newcommand{\ASLC}{\mbox{ASLC}}
\newcommand{\SQSD}{\mbox{SQSD}}
\newcommand{\SQDIM}{\mbox{SQ-DIM}}
\newcommand{\SQSDIM}{\mbox{SQ-SDIM}}
\newcommand{\N}{{\mathcal N}}

\newcommand{\eg}{e.g.\ }

\newcommand{\cf}{{\em cf.\ }}
\newcommand{\fr}[1]{\frac{1}{#1}}

\newcommand{\A}{{\mathcal A}}

\newcommand{\B}{{\mathcal B}}

\newcommand{\C}{{\mathcal C}}
\newcommand{\D}{{\mathcal D}}
\newcommand{\F}{{\mathcal F}}

\newcommand{\R}{{\mathbb R}}
\newcommand{\E}{\mathop{\mathbf E}}

\newcommand{\pr}{\mathop{\mathbf{Pr}}}
\newcommand{\ra}{\rangle}
\newcommand{\la}{\langle}
\newcommand{\cond}{\ |\ }

\newcommand{\zon}{\{0,1\}^n}

\newcommand{\etal}{{\em et al.\ }}

\newcommand{\pmi}{\{-1,1\}}

\newcommand{\ti} \tilde

\newcommand{\eps}{\epsilon}

\newcommand{\exoracle}[2]{\mbox{EX}(#2,#1)}

\newcommand{\sign}[1]{\mbox{\tt{sign}}(#1)}

\newcommand{\equ}[1]{

\begin{equation}
#1
\end{equation}}

\newcommand{\ignore}[1]{\relax}

\newcommand{\alequn}[1]{\begin{align*} #1 \end{align*}}

\newcommand{\eat}[1]{}

\newtheorem{theorem}{Theorem}[section]
\newtheorem{lemma}[theorem]{Lemma}
\newtheorem{claim}[theorem]{Claim}

\newtheorem{remark}[theorem]{Remark}

\newtheorem{definition}[theorem]{Definition}

\newenvironment{proof}{\noindent \textbf{Proof:}}{\hfill{$\Box$}}

\title{A Complete Characterization of Statistical Query Learning with Applications to Evolvability\footnote{Earlier version of this work appeared in the proceedings of the 44th IEEE Symposium on Foundations of Computer Science, 2009.}}
\author{Vitaly Feldman \\
{\em IBM Almaden Research Center}\\
\tt{vitaly@post.harvard.edu}}
\date{}
\begin{document}

\maketitle
\abstract{
Statistical query (SQ) learning model of Kearns is a natural restriction of the PAC learning model in which a learning algorithm is allowed to obtain estimates of statistical properties of the examples but cannot see the examples themselves \cite{Kearns:98}. We describe a new and simple characterization of the query complexity of learning in the SQ learning model. Unlike the previously known bounds on SQ learning \cite{BlumFJ+:94,BshoutyFeldman:02,Yang:05,BalcazarCGKL:07,Simon:07} our characterization preserves the accuracy and the efficiency of learning. The preservation of accuracy implies that our characterization gives the first characterization of SQ learning in the agnostic learning framework of Haussler \cite{Haussler:92}, and Kearns, Schapire and Sellie \cite{KearnsSS:94}. The preservation of efficiency is achieved using a new boosting technique and allows us to derive a new approach to the design of evolution algorithms in Valiant's model of evolvability \cite{Valiant:09}. We use this approach to demonstrate the existence of a large class of monotone evolution algorithms based on square loss performance estimation. These results differ significantly from the few known evolution algorithms and give evidence that evolvability in Valiant's model is a more versatile phenomenon than there had been previous reason to suspect.
}

\section{Introduction}
We study the complexity of learning in Kearns' well-known {\em statistical query} (SQ) learning model \cite{Kearns:98}. Statistical query learning is a natural restriction of the PAC learning model in which a learning algorithm is allowed to obtain estimates of statistical properties of the examples but cannot see the examples themselves. Formally, the learning algorithm is given access to STAT($f,D$) -- a {\em statistical query oracle} for the unknown target function $f$ and distribution $D$ over some domain $X$. A query to this oracle is a function of an example $\phi : X \times \pmi \rightarrow \pmi$. The oracle may respond to the query with any value $v$ satisfying $|\E_{x\sim D}[\phi(x,f(x))] - v| \leq \tau$ where $\tau \in [0,1]$ is the {\em tolerance} of the query.

Kearns demonstrated that any learning algorithm that is based on statistical queries can be automatically converted to a learning algorithm robust to random classification noise of arbitrary rate smaller than the information-theoretic barrier of $1/2$ \cite{Kearns:98}. Most known learning algorithms can be converted to statistical query algorithms and hence the SQ model proved to be a powerful technique for the design of noise-tolerant learning algorithms (\eg  \cite{Kearns:98,Bylander:94,BlumFKV:97,DunaganVempala:04}). In fact, since the introduction of the model virtually all\footnote{A notable exception is the algorithm for learning parities of Blum \etal \cite{BlumKW:03} which is tolerant to random noise, albeit not in the same strong sense as the algorithms derived from SQs.} known noise-tolerant learning algorithms were obtained from SQ algorithms. The basic approach was also extended to deal with noise in numerous other learning scenarios and has also found applications in several other areas including privacy-preserving learning and learning on multi-core systems \cite{BansalBC:02,BlumDMN:05,ChuKLYBNO:06,KLNRS:08}. This makes the study of the complexity of SQ learning crucial for the understanding of noise-tolerant learning and PAC learning in general.

Kearns has also demonstrated that there are information-theoretic impediments unique to SQ learning: parity functions require an exponential number of SQs to be learned \cite{Kearns:98}. Further, Blum \etal proved that the number of SQs required for weak learning (that is, one that gives a non-negligible advantage over the random guessing) of a concept class $C$ is characterized by a relatively simple combinatorial parameter of $C$ called the {\em statistical query dimension} $\SQDIM(C,D)$ \cite{BlumFJ+:94}. $\SQDIM(C,D)$ measures the maximum number of ``nearly uncorrelated" (relative to distribution $D$) functions in $C$. Bshouty and Feldman gave an alternative way to characterize weak learning by statistical query algorithms that is based on the number of functions required to weakly approximate each function in $C$ \cite{BshoutyFeldman:02}. These bounds for weak learning were strengthened and extended to other variants of statistical queries in several works \cite{BlumKW:03,Yang:05,Feldman:08ev}. Notable applications of these bounds are lower bounds on SQ-DIM of several concept classes by Klivans and Sherstov \cite{KlivansSherstov:07a} and an upper-bound on the SQ dimension of halfspaces by Sherstov \cite{Sherstov:07a}.

While the query complexity of weak SQ learning is fairly well-studied, few works have addressed the query complexity of strong SQ learning. It is easy to see that there exist classes of functions for which strong SQ complexity is exponentially higher than the weak SQ complexity. One such example is learning of monotone functions with respect to the uniform distribution. The complexity of weak SQ learning and hence the statistical query dimension are polynomial \cite{KearnsLV:94,BshoutyTamon:96}. However, strong PAC learning of monotone functions with respect to the uniform distribution requires an exponential number of examples and hence an exponential number of statistical queries \cite{KearnsLV:94,BlumBL:98}. In addition, it is important to note that the statistical query dimension and other known notions of statistical query complexity are distribution-specific and therefore one cannot directly invoke the equivalence of weak and strong SQ learning in the distribution-independent setting \cite{AslamDecatur:98b}.
The first explicit\footnote{An earlier work has also considered this question but the characterization that was obtained is in terms of query-answering protocols that are essentially specifications of non-adaptive algorithms \cite{BalcazarCGKL:07}.} characterization of strong SQ learning with respect to a fixed distribution $D$ was only recently derived by Simon \cite{Simon:07}.

\subsection{Our Results}
Our main result is a complete characterization of the query complexity of SQ learning in both the PAC and the agnostic models. Informally, our characterization states that a concept class $C$ is SQ learnable over a distribution $D$ if and only if for every real-valued function $\psi$, there exists a small (i.e. polynomial-size) set of functions $G_\psi$ such that for every $f \in C$, if $\sign{\psi}$ is not ``close" to $f$ then one of the functions in $G_\psi$ is ``noticeably" correlated with $f - \psi$. More formally, for a distribution $D$ over $X$, we define the (semi-)inner product over the space of real-valued functions on $X$ as $\la \phi, \psi \ra_D = \E_{x \sim D}[\phi(x) \cdot \psi(x)]$. Then $C$ is SQ learnable to accuracy $\eps$ if and only if for every  $\psi: X \rightarrow [-1,1]$,  there exists a set of functions $G_\psi$ such that (1) for every $f \in C$, if $\pr_D[\sign{\psi} \neq f] \geq \eps$ then $|\la g, f - \psi \ra_D | \geq \gamma$ for some $g \in G_\psi$; (2) $|G_\psi|$ is polynomial and $\gamma >0 $ is inverse-polynomial in $1/\eps$ and $n$ (the size of the learning problem). We refer to this characterization as {\em approximation-based}.

For Boolean functions Bshouty and Feldman proved that the number of functions required to weakly approximate every function in a set of functions $C$ is polynomially related to the (weak) statistical query dimension of $C$ \cite{BshoutyFeldman:02}. We use a generalization of this idea to real-valued functions to obtain another characterization of SQ learnability. Specifically, for a set of functions $F$, we say that $\SQDIM(F,D)$ equals $d$ if $d$ is the largest number for which there are $d$ functions $f_1,f_2,\ldots, f_d \in F$,  such that for every $i\neq j$, $ |\la f_i,f_j \ra_D| \leq 1/d$. Our approximation-based characterization leads to the following characterization based on $\SQDIM$: $\SQSDIM(C,D,\eps) = \sup_{\psi} \{\SQDIM((C \setminus B^D(\sign{\psi},\eps))- \psi,D)\}$, where $B^D(\sign{\psi},\eps)$ is the set of functions that differ from $\sign{\psi}$ on at most $\eps$ fraction of $X$ and $F-\psi = \{f-\psi \cond  f \in F\}$. When the correlation between functions is interpreted as an inner product, $\SQDIM(F,D)$ measures the largest number of almost orthogonal (relative to $D$) functions in $F$. Therefore we refer to this characterization as {\em orthogonality-based}.

An important property of both of these characterizations is that the accuracy parameter in the dimension corresponds to the accuracy parameter $\eps$ of learning (up to the tolerance of the SQ learning algorithm). The advantage of the approximation-based characterization is that it preserves computational efficiency of learning. Namely, the set of approximating functions for $\eps$-accurate learning can be computed efficiently if and only if there exists an efficient SQ learning algorithm achieving error of at most $\eps$. The orthogonality-based characterization does not preserve efficiency but is more easy to analyze when proving lower bounds. Neither of these properties are possessed by the previous characterizations of strong SQ learning \cite{BalcazarCGKL:07,Simon:07,Szorenyi:09}.

The preservation of accuracy implies that both of our characterizations can be naturally extended to agnostic learning by replacing the concept class $C$ with the set of all functions that are $\Delta$-close to at least one concept in $C$ (see Th.~\ref{th:learn2ASQSD}). Learning in this model is notoriously hard and this is readily confirmed by the SQ dimension we introduce. For example, in Theorem \ref{th:agn-conjunction-lb} we prove that the SQ dimension of agnostic learning of monotone conjunctions with respect to the uniform distribution is super-polynomial. This provides new evidence that agnostic learning of conjunctions is a hard problem even when restricted to the monotone case over the uniform distribution. The preservation of accuracy is critical for the generalization to agnostic learning since, unlike in the PAC model, achieving, for example, twice the error (i.e. $2 \cdot \Delta$) might be a substantially easier task than learning to accuracy $\Delta + \eps$ (for example when $\Delta \geq 1/4$).

We note that the characterization of (strong) SQ learning by Simon \cite{Simon:07} has some similarity to ours. It also examines weak statistical query dimension of $F - \psi$ for $F \subseteq C$ and some function $\psi$. However, the maximization is over all sets of functions $F$ satisfying several properties and $\phi$ is fixed to be the average of functions in $F$. Simon's SQ dimension and the characterization were substantially simplified in a very recent and independent work of Sz\"{o}r\'{e}nyi \cite{Szorenyi:09}. His elegant characterization result is based on measuring the maximum number of functions in $C$ whose pairwise correlations are nearly identical. It was shown by Sz\"{o}r\'{e}nyi that his dimension can be directly related to $\SQSDIM$. His proof of the upper bound on the SQ learning complexity uses an inefficient algorithm and therefore his characterization does not preserve efficiency of computation. The proof of the lower bound doubles the accuracy (that is, the dimension with accuracy $\eps$ lower bounds the SQ learning complexity with accuracy $2\eps$). Therefore the lower bound does not preserve the accuracy of learning. The techniques in his proofs are not directly comparable to ours.


\subsection{Overview of the Proof}
To prove the first direction of our characterization we simulate the SQ learning algorithm for $C$ while replying to its statistical queries using $\psi$ in place of the unknown target function $f$. If $\psi$ is not close to $f$ then one of the queries in this execution has to distinguish between $f$ and $\psi$, giving a function that weakly approximates $f-\psi$. Hence the polynomial number of queries in this execution implies the existence of the set $G_\psi$ with the desired property.

For the second direction we use the fact that $\la g, f - \psi \ra_D  \geq \gamma$ means that $g$ ``points" in the direction of $f$ from $\psi$,
that is, $\psi + \gamma \cdot g$ is closer to $f$ than $\psi$ by at least $\gamma^2$ in the norm corresponding to our inner product. Therefore one can ``learn" the target function $f$ by taking steps in the direction of $f$ until the hypothesis converges to $f$. This argument requires the hypothesis at each step to have range in $[-1,1]$ and therefore we apply a projection step after each update. This process is closely related to projected gradient descent -- a well-known technique in a number of areas. The closest analogues of this technique in learning are some boosting algorithms (\eg \cite{BarakHK:09}). In particular, our algorithm is closely related to the hard-core set construction of Impagliazzo \cite{Impagliazzo:95} adapted to boosting by Klivans and Servedio \cite{KlivansServedio:03}. The proof of our result can also be seen as a new type of boosting algorithm that instead of using a weak learning algorithm on different distributions uses a weak learning algorithm on different target functions (namely $f - \psi$). This connection is explored in \cite{Feldman:10ab}.

\subsection{Applications to Evolvability}
The characterization and its efficiency-preserving proofs imply that if $C$ is SQ learnable then for every hypothesis function $\psi$, there exists a small and {\em efficiently computable}  set of functions $N(\psi)$ such that if $\psi$ is not ``close" to $f \in C$ then one of the functions in $N(\psi)$ is ``closer" to $f$ than $\psi$ (Th.~\ref{th:sqlearn2neighbor}). This property implies that every SQ learnable $C$ is learnable by a canonical learning algorithm which learns $C$ via a sequential process in which at every step the best hypothesis is chosen from a small and fixed pool of hypotheses ``adjacent" to the current hypothesis. This type of learning has been recently proposed by Valiant as one that can explain the acquisition of complex functionality by living organisms through the process of evolution guided by natural selection \cite{Valiant:09}.
One particular important issue raised by the model is the ability of an evolution algorithm to converge to a high accuracy hypothesis without relying on decreases in the performance in the process of evolving. We refer to this property as being {\em monotone}. Monotonicity allows an evolution algorithm to adjust to a change of the target function without sacrificing the performance of the current hypothesis. Existence of evolution algorithms that are robust to such changes could explain the ability of some organisms to adapt to changes in environmental conditions without the need for a ``restart". Monotonicity is not required in the basic Valiant's model and the power of evolvability without this requirement was resolved in our recent work \cite{Feldman:08ev,Feldman:09robust}. There we showed that, depending on how the performance of hypotheses is measured, evolvability is equivalent to either the SQ learnability or the learnability by restricted SQs referred to as correlational SQs (see Sec.~\ref{sec:define-sq} for the definition). Prior to this work monotone evolvability was only known for several very restricted classes of functions and distributions,  namely, conjunctions over the uniform distribution \cite{Valiant:09}\footnote{Monotonicity was demonstrated explicitly by Kanade \etal \cite{KanadeVV:10}.}, decision lists over the uniform distribution \cite{Michael:07}, and the singletons (functions that are positive on a single point) over all distributions \cite{Feldman:09robust}. Interestingly, there are no known non-monotone evolution algorithms which were designed for specific concept classes (rather than obtained through general transformation from SQ learning algorithms). Valiant's original model and the results in \cite{Valiant:09} and \cite{Feldman:08ev} use Boolean hypotheses and the correlation (or, equivalently, the probability of agreement) is used to measure the performance of hypotheses. In Michael's work measuring performance using the quadratic loss over all real-valued hypotheses was introduced and used to prove evolvability of decision lists \cite{Michael:07}. The power of using different loss functions over real-valued hypotheses was studied in \cite{Feldman:09robust} where we showed that evolvability with the Boolean loss implies evolvability with the quadratic loss (and all other loss functions) but not vice versa.

Our canonical learning algorithms can be fairly easily translated into evolution algorithms demonstrating that every concept class $C$ SQ-learnable with respect to a distribution $D$, is evolvable monotonically over $D$ (Th.~\ref{th:evolve-LSQ}) when the performance is measured using the quadratic loss. While we do not know how to extend this general method to the more robust distribution-independent evolvability, we show that the underlying ideas can be useful for this purpose as well. Namely, we prove distribution-independent and monotone evolvability of Boolean disjunctions (or conjunctions) using a simple and natural mutation algorithm (Th.~\ref{th:evolve-disj}). The mutation algorithm is based on slight adjustments of the contribution of each of the Boolean variables while bounding the total value of contributions (which corresponds to the projection step). 

The stronger properties of the quadratic loss function on real-valued hypotheses were first exploited in Michael's algorithm for evolving decision lists \cite{Michael:07}. The model in that work is slightly different from ours as it uses representations of unbounded range (versus the $[-1,1]$ range in our work) and a scaled quadratic loss function (with the scale determined by the desired accuracy of the evolution algorithm). Hence the result in \cite{Michael:07} will not hold in the model we consider here (which was defined in \cite{Feldman:09robust}). The analysis in his work relies heavily on the particular properties the Fourier transform of decision lists when learned over the uniform distribution and is not directly related to the broad setting we consider here. Formal definitions of the model and the results are given in Section \ref{sec:apply-evolve}.

\subsection{Relation to the Earlier Version}
Since the appearance of the earlier version of this work \cite{Feldman:09sqd} we have found ways to strengthen some of the parameters of the characterizations. As a result the dimensions used here differ from the ones introduced in \cite{Feldman:09sqd}. Also, unlike the dimension we use here, the SQD$_\eps$ dimension in \cite{Feldman:09sqd} preserves the output hypothesis space and hence is suitable for characterizing {\em proper} learning. To emphasize the difference we use different notation for the dimensions defined in the two versions of the work. In addition, the characterization of learning in the agnostic model is now simplified using recent distribution-specific agnostic boosting algorithms \cite{Feldman:10ab,KalaiKanade:09}.

\section{Preliminaries}
\label{sec:general-defs}
For a positive integer $\ell$, let $[\ell]$ denote the set $\{1,2,\ldots,\ell\}$. We denote the domain of our learning problems by $X$ and let $\F^\infty_1$ denote the set of all functions from $X$ to $[-1,1]$ (that is all the functions with $L_\infty$ norm bounded by 1). It will be convenient to view a distribution $D$ over $X$ as defining the product $\la \phi, \psi \ra_D = \E_{D}[\phi(x) \cdot \psi(x)]$ over the space of real-valued functions on $X$. It is easy to see that this is simply a non-negatively weighted version of the standard dot product over $\R^X$ and hence is a positive semi-inner product over $\R^X$. The corresponding norm is defined as $\| \phi \|_D = \sqrt{\E_D[\phi^2(x)]} = \sqrt{\la \phi, \phi \ra_D}$. We define an $\eps$-ball around a Boolean function $h$ as $B^D(h,\eps) = \{g: X \rightarrow \pmi \cond \pr_D[f \neq g] \leq \eps \}$. For two real-valued functions $\phi$ and $\psi$ we let $L_1^D(\phi,\psi) = \E_D[|\phi(x) - \psi(x)|]$. For a set of real-valued functions $F$ and a real-valued function $\psi$ we denote by $F - \psi = \{f - \psi \cond f \in F \}$. For a real value $a$, we denote its projection to $[-1,1]$ by $P_1(a)$. That is, $P_1(a) = a$ if $|a| \leq 1$ and $P_1(a) = \sign{a}$, otherwise.

\subsection{PAC Learning}
\label{sec:define-PAC}
For a {\em domain} $X$, a {\em concept class} over $X$ is a set of $\pmi$-valued functions over $X$ referred to as {\em concepts}. A concept class together with a specific way to represent all the functions in the concept class is referred to as a {\em representation class}. For brevity, we often refer to a representation class as just a concept class with some implicit representation scheme.

There is often a complexity parameter $n$ associated with the domain $X$ and the concept class $C$ such as the number of Boolean variables describing an element in $X$ or the number of real dimensions. In such a case it is understood that $X =\bigcup_{n\geq 1} X_n$ and $C =\bigcup_{n\geq 1} C_n$. We drop the subscript $n$ when it is clear from the context. In some cases it useful to consider another complexity parameter associated with $C$: the minimum description length of $f$ under the representation scheme of $C$. Here, for brevity, we assume that $n$ (or a fixed polynomial in $n$) bounds the description length of all functions in $C_n$.

The models we consider are based on the well-known PAC learning model introduced by Valiant \cite{Valiant:84}. Let $C$ be a representation class over $X$. In the basic PAC model a learning algorithm is given examples of an unknown function $f$ from $C$ on points randomly chosen from some unknown distribution $D$ over $X$ and should produce a hypothesis $h$ that approximates $f$. Formally, an {\em example oracle} $\exoracle{D}{f}$ is an oracle that upon being invoked returns an example $\langle x,f(x)\rangle$, where $x$ is chosen randomly with respect to $D$, independently of any previous examples.

An algorithm is said to PAC learn $C$ in time $t$ if for every $\epsilon > 0$, $f \in C$, and distribution $D$ over $X$, the algorithm given $\epsilon$ and access to $\exoracle{D}{f}$ outputs, in time $t$ and with probability at least $2/3$, a hypothesis $h$ that is evaluatable in time $t$ and satisfies $\pr_{D}[f(x) \neq h(x)] \leq \epsilon$. For convenience we also allow real-valued hypotheses in $\funcs$. Such a hypothesis needs to satisfy $\la f(x), h(x) \ra_D \geq 1-2\epsilon$. A real-valued hypothesis $\phi(x)$ can be also thought of as a randomized Boolean hypothesis $\Phi(x)$, such that $\phi(x)$ equals the expected value of $\Phi(x)$. Hence $\la f(x), \phi(x) \ra_D \geq 1-2\epsilon$ is equivalent to saying that the expected error of $\Phi(x)$ is at most $\eps$. We say that an algorithm {\em efficiently} learns $C$  when $t$ is upper bounded by a polynomial in $n$, $1/\epsilon$.

The basic PAC model is also referred to as {\em distribution-independent} learning to distinguish it from {\em distribution-specific} PAC learning in which the learning algorithm is required to learn only with respect to a single distribution $D$ known in advance.

A {\em weak} learning algorithm \cite{KearnsValiant:94} is a learning algorithm that produces a hypothesis whose disagreement with the target concept is noticeably less than $1/2$ (and not necessarily less than any $\epsilon > 0$).  More precisely, a weak learning algorithm produces a hypothesis $h \in \funcs$ such that $\la f(x), h(x) \ra_D \geq 1/p(n)$ for some fixed polynomial $p$.

\subsection{Agnostic Learning}
The {\em agnostic} learning model was introduced by Haussler \cite{Haussler:92} and Kearns \etal \cite{KearnsSS:94} in order to model situations in which the assumption that examples are labeled by some $f \in \C$ does not hold. In the most general version of the model the examples are generated from some unknown distribution $A$ over $X \times \pmi$. The goal of an agnostic learning algorithm for a concept class $C$ is to produce a hypothesis whose error on examples generated from $A$ is close to the best possible by a concept from $C$.  Any distribution $A$ over $X \times \pmi$ can be described uniquely by its marginal distribution $D$ over $X$ and the expectation of the label $b$ given $x$. That is, we refer to a distribution $A$ over $X \times \pmi$ by a pair $(D_A,\phi_A)$ where $D_A(z) = \pr_{\la x,b\ra\sim A}[x=z]$ and $$\phi_A(z) = \E_{\la x,b\ra\sim A}[b \cond z=x].$$

Formally, for a function $h \in \funcs$ and a distribution $A=(D,\phi)$ over $X \times \pmi$, we define $$\Delta(A,h) = L_1^D(\phi,h)/2\ .$$ Note that for a Boolean function $h$, $\Delta(A,h)$ is exactly the error of $h$ in predicting an example drawn randomly from $A$ or $\pr_{\la x,b \ra \sim A}[h(x) \neq b]$. For a concept class $C$, let $\Delta(A,C) = \inf_{h \in  C}\{\Delta(A,h)\}\ .$

Kearns \etal \cite{KearnsSS:94} define agnostic learning as follows.
\begin{definition}
\label{def:agnostic-model}
An algorithm $\A$ {\em agnostically} learns a representation class $C$ if for every $\epsilon > 0$, distribution $A$ over $X \times \pmi$, $\A$ given access to examples drawn randomly from $A$, outputs, with probability at least $2/3$, a hypothesis $h \in \funcs$ such that $\Delta(A,h) \leq \Delta(A,C) + \eps$.
\end{definition}
As in the PAC learning, the learning algorithm is {\em efficient} if it runs in time polynomial $1/\eps$ and $n$.

More generally,  for $0 < \alpha \leq \beta   \leq 1/2$ an $(\alpha,\beta)$-agnostic learning algorithm is the algorithm that produces a hypothesis $h$ such that $\Delta(A,h) \leq \beta$ whenever $\Delta(A,C) \leq \alpha$.
In the distribution-specific version of this model, learning is only required for every $A = (D,\phi)$, where $D$ equals to some fixed distribution known in advance.

\subsection{The Statistical Query Learning Model}
\label{sec:define-sq}
In the {\em statistical query model} of Kearns \cite{Kearns:98} the learning algorithm is given access to
STAT($f,D$) -- a {\em statistical query oracle} for target concept
$f$ with respect to distribution $D$ instead of $\exoracle{D}{f}$. A query to this oracle is a pair $(\psi,\tau)$ where $\psi : X \times \pmi \rightarrow \pmi$ and $\tau >0$. The oracle may respond to the query with any value $v$ satisfying $|\E_D[\psi(x,f(x))] - v| \leq \tau$  where $\tau$ is referred to as the {\em tolerance} of the query. For convenience, we allow the query functions to be real-valued in the range $[-1,1]$. As it has been observed by Aslam and Decatur \cite{AslamDecatur:98}, this extension is equivalent to the original SQ model.

An algorithm $\A$ is said to learn $C$ in time $t$ from statistical queries of tolerance $\tau$
if $\A$ PAC learns $C$ using STAT($f,D$) in place of the example oracle. In addition, each query $\psi$ made by $\A$ has tolerance $\tau$ and can be evaluated in time $t$. The {\em statistical query learning complexity} of $C$ over $D$ is the minimum number of queries of tolerance $\tau$ sufficient to learn $C$ over $D$ to accuracy $\eps$ and is denoted by $\SLC(C,D,\eps,\tau)$.

The algorithm is said to (efficiently) SQ learn $C$ if $t$ is polynomial in $n$ and $1/\epsilon$, and  $\tau$ is lower-bounded by the inverse of a polynomial in $n$ and $1/\epsilon$.

The SQ learning model extends to the agnostic setting analogously. That is, random examples from $A$ are replaced by queries to the SQ oracle STAT$(A)$. For a query $\psi$ as above, STAT$(A)$ returns a value $v$ satisfying $|\E_{\la x,b\ra\sim A}[\psi(x,b)] - v| \leq \tau$. We denote the {\em agnostic statistical query learning complexity} of $C$ over $D$ by $\ASLC(C,D,\eps,\tau)$.

A {\em correlational} statistical query is a statistical query for a correlation of a function over $X$ with the target \cite{BshoutyFeldman:02}. Namely the query function $\psi(x, \ell) \equiv \phi(x) \cdot \ell$ for a function $\phi \in \funcs$. We say that a query is {\em target-independent} if  $\psi(x,\ell) \equiv \phi(x)$ for a function $\phi \in \funcs$, that is, if $\psi$ is a function of the point $x$ alone. We will need the following simple fact by Bshouty and Feldman \cite{BshoutyFeldman:02} to relate learning by statistical queries to learning by CSQs.
  \begin{lemma}[\cite{BshoutyFeldman:02}]
  \label{lem:sq-simple}
  For any function $\psi: X \times \pmi \rightarrow [-1,1]$,  $\psi(x,\ell) \equiv \phi_1(x) \cdot \ell + \phi_2(x)$, for some $\phi_1,\phi_2 \in \funcs$. In particular a statistical query $(\psi,\tau)$ with respect to any distribution $D$ can be answered using a correlational statistical query $(\phi_1(x) \cdot \ell,\tau_1)$ and a target-independent query $(\phi_2(x),\tau_2)$, for any $\tau_1,\tau_2$ such that $\tau = \tau_1+\tau_2$.
  \end{lemma}


\subsection{ (Weak) SQ Dimension}
\label{sec:weaksqd}
Blum \etal showed that concept classes weakly SQ learnable using only a polynomial number of statistical queries of inverse polynomial tolerance are exactly the concept classes that have polynomial {\em statistical query dimension} or SQ-DIM \cite{BlumFJ+:94}. The dimension is based on the largest number of almost orthogonal (using the $\la\cdot,\cdot\ra_D$ inner product) functions in the set.
\begin{definition}[\cite{BlumFJ+:94,Yang:05}]
\label{def:old-weak-sqd}
For a concept class $C$  we say that $\SQDIM(C,D) = d$ if $d$ is the largest value for which there exist $d$ functions $f_1, f_2, \ldots, f_d \in C$ such that for every $i\neq j$, $|\la f_i , f_j \ra_D| \leq 1/d$.
\end{definition}
Bshouty and Feldman gave an alternative way to characterize weak learning by statistical query algorithms that is based on the number of functions required to weakly approximate each function in the set \cite{BshoutyFeldman:02}.
\begin{definition}
\label{def:weak-sqd}
For a concept class $C$ and $\gamma > 0$ we say that $\SQD(C,D,\gamma) = d$ if there exists a set of $d$ functions $G \subset \funcs$ such that for every $f\in C$,  $|\la f , g \ra_D| \geq \gamma$ for some $g \in G$. In addition, no value smaller than $d$ has this property.
\end{definition}
Bshouty and Feldman show that a concept class $C$ is weakly SQ learnable over $D$ using a polynomial number of queries if and only if $\SQD(C,D,1/t(n)) = d(n)$ for some polynomials $d(\cdot)$ and $t(\cdot)$ \cite{BshoutyFeldman:02}.
It is also possible to relate SQD and SQ-DIM more directly. It is well-known that the maximal set of almost orthogonal functions in $C$ is also the approximating set for $C$. In other words, $\SQD(C,D,1/d) \leq d$, where $d = \SQDIM(C,D)$. The connection in the other direction is implicit in the work of Blum \etal \cite{BlumFJ+:94}. Here we will use a stronger version given by Yang \cite{Yang:05} (see \cite{Szorenyi:09} for a recent simpler proof).
\begin{lemma}[\cite{Yang:05}]
\label{lem:SQD2SSQDIM}
Let $C$ be a concept class and $D$ be a distribution over $X$. Then
$\SQD(C,D,d^{-1/3}) \geq d^{1/3}/2$, where $d=\SQDIM(C,D)$.
\end{lemma}

\section{Strong SQ Dimension}
\label{sec:strong-sqd}
In this section we give a generalization of the weak statistical query dimension to strong learning. We first extend the approximation-based characterization of Bshouty and Feldman \cite{BshoutyFeldman:02} and then obtain an orthogonality-based characterization from it.
\subsection{Approximation-Based Characterization}
In order to define our strong statistical query dimension we first need to generalize the approximation-based characterization of Bshouty and Feldman \cite{BshoutyFeldman:02} to sets of real-valued functions rather than just concept classes. To achieve this we simply note that the definition of $\SQD(C,D,\gamma)$ does not use the fact that functions in $C$ are Boolean and hence we can define $\SQD(F,D,\gamma)$ for any set of real-valued functions $F$ in exactly the same way. We now define the strong statistical query dimension of a class of functions $C$.
\begin{definition}
\label{def:ssqd}
For a concept class $C$, distribution $D$ and $\eps,\gamma > 0$ we define
$$\SQSD(C,D,\eps,\gamma) = \sup_{\psi \in \funcs}\left\{\SQD(C\setminus B^D(\sign{\psi},\eps)- \psi,D,\gamma) \right\},$$
In other words, we say that $\SQSD(C,D,\eps,\gamma) = d$ if for every $\psi \in \funcs$, there exists a set of $d$ functions $G_\psi \subset \funcs$ such that for every $f \in C$, either
\begin{enumerate}
\item $\pr_D[f(x) \neq \sign{\psi(x)}] \leq \eps$ or 
\item there exists $g \in G_\psi$ such that $|\la f - \psi , g \ra_D| \geq \gamma$.
\end{enumerate}
In addition, no value smaller than $d$ has this property.
\end{definition}

We now give a simple proof that $\SQSD(C,D,\eps,\gamma)$ characterizes (within a polynomial) the number of statistical queries required to learn $C$ over $D$ with accuracy $\eps$ and query tolerance $\gamma$.
\begin{theorem}
\label{th:learn2SQSD}
For every concept class $C$, distribution $D$ over $X$ and $\eps,\tau > 0$,
$$\SLC(C,D,\eps,\tau) \geq \SQSD(C,D,\eps+\tau,\tau)-2\ .$$
\end{theorem}
\begin{proof}
Let $\A$ be a SQ algorithm that learns $C$ over $D$ using $q= SLC(C,D,\eps,\tau)$ queries of tolerance $\tau$. According to Lemma \ref{lem:sq-simple}, we can decompose every SQ of $\A$ into a correlational and a target-independent queries. The distribution $D$ is fixed and therefore any target-independent query of $\A$ for function $\phi(x)$ can always be answered with the exact value $\E_D[\phi(x)]$, in other words with tolerance $0$. Therefore it is sufficient to answer the $q$ correlational SQs of $\A$ with tolerance $\tau$.

Now let $\psi \in \funcs$ be any function. The set $G_\psi$ is constructed as follows. Simulate algorithm $\A$ and for every correlational query $(\phi_i \cdot \ell,\tau)$ add $\phi_i$ to $G_\psi$ and respond to the query with the value $\la \psi , \phi_i \ra_D = \E_D[\phi_i(x)\cdot \psi(x)]$. Continue the simulation until $\A$ outputs a hypothesis $h_\psi$. Add $\sign{\psi}$ and $h_\psi$ to $G_\psi$.

First, by the definition of $G_\psi$, $q \geq |G_\psi| - 2$. Now, let $f$ be any function in $C$. If there does not exist $g \in G_\psi$ such that $|\la f - \psi , g \ra_D| \geq \tau$ then for every correlational query function $\phi_i \in G_\psi$, $|\la \psi , \phi_i \ra_D - \la f , \phi_i \ra_D| < \tau \ .$ This means that in our simulation, $\la \psi , \phi_i \ra_D$ is within $\tau$ of $\la f , \phi_i \ra_D$. Therefore the answers provided by our simulator are valid for the execution of $\A$ when the target function is $f$. That is they could have been returned by STAT$(f,D)$ with tolerance $\tau$. Therefore, by the definition of $\A$, the
hypothesis $h_\psi$ satisfies $\la f, h_\psi \ra_D \geq 1-2\eps$. Both $\sign{\psi}$ and $h_\psi$ are in $G_\psi$ and therefore we also know that
$|\la f - \psi , \sign{\psi} \ra_D| \leq \tau$ and $|\la f - \psi , h_\psi \ra_D| \leq \tau$. These conditions imply that $\la f, \sign{\psi} \ra_D \geq \la \psi, \sign{\psi} \ra_D - \tau$ and $\la \psi, h_\psi \ra_D \geq \la f, h_\psi \ra_D - \tau$. In addition for all $\psi,h_\psi \in \funcs$, $\la \psi, \sign{\psi} \ra_D  \geq  \la \psi, h_\psi \ra_D$. By combining these inequalities, we
 conclude that
$$ \la f, \sign{\psi} \ra_D \geq \la \psi, \sign{\psi} \ra_D  - \tau \geq  \la \psi, h_\psi \ra_D - \tau \geq  \la f, h_\psi \ra_D - 2\tau \geq 1-2\eps-2\tau\ ,$$ which is equivalent to $\pr_D[f(x) \neq \sign{\psi(x)}] \leq \eps +\tau$. In other words, if there does not exist $g \in G_\psi$ such that $|\la f - \psi , g \ra_D| \geq \tau$ then $f \in B^D(\sign{\psi},\eps +\tau)$, giving us the claimed inequality.
\end{proof}
\begin{remark}
If $\A$ is randomized then it can be converted to a non-uniform deterministic algorithm (in the sense of having access to a fixed polynomial size {\em advice} string) via a standard confidence boosting transformation (\eg \cite{BshoutyFeldman:02}). This transformation increases the number of queries by a polynomial factor but leaves the accuracy of learning and the tolerance of queries unchanged. Therefore, up to a polynomial factor, Theorem \ref{th:learn2SQSD} also applies to SQ learning by randomized algorithms.
\end{remark}
We now establish the other direction of our characterization.


\begin{theorem}
\label{th:ssq2alg}
For every concept class $C$, distribution $D$ over $X$ and $\eps,\tau > 0$,
$$\SLC(C,D,\eps,\tau) \leq \SQSD(C,D,\eps,4 \cdot \tau)/(3\tau^2)\ .$$
\end{theorem}
\begin{proof}
Let $d = \SQSD(C,D,\eps,4 \cdot \tau)$. Our learning algorithm for $C$ builds an approximation to the target function $f$ in steps. In each step we have a current hypothesis $\psi_i \in \funcs$. If $\sign{\psi_i}$ is not $\eps$-close to $f$ then we find a function $g \in G_{\psi_i}$ such that $|\la f - \psi_i, g \ra_D | \geq \gamma$. Such $g$ can be viewed as a vector ``pointing" in the direction of $f$ from $\psi_i$. We therefore set $\psi'_{i+1} = \psi _i + \la f - \psi_i, g \ra_D \cdot g$. As we will show $\psi'_{i+1}$ is closer (in distance measured by $\|\cdot\|_D$) to $f$ than $\psi_i$. However $\psi'_{i+1}$ is not necessarily in $\funcs$. We define $\psi_{i+1}$ to be the projection of $\psi'_{i+1}$ onto $\funcs$. As we will show this projection step only decreases the distance to the target function. We will now provide the details of the proof.

Let $\psi_0 \equiv 0$. Given $\psi_i$ we define $\psi_{i+1}$ as follows. Let $G_{\psi_i}$ be the set of size at most $d$ that correlates with every function in $C\setminus B^D(\sign{\psi_i},\eps) - \psi_i$ (as given by Definition \ref{def:ssqd}). For every $g \in G_{\psi_i}$ we make a query for $\la f, g \ra_D$ to STAT$(f,D)$ with tolerance $\tau$ and denote the answer by $v(g)$. If there exists $g \in G_{\psi_i}$ such that $|v(g) - \la \psi_i, g \ra_D | \geq 3\tau$ then we set $g_i = g$, $\gamma_i = v(g_i) - \la \psi_i, g_i \ra_D$, and  $\psi'_{i+1} = \psi_i + \gamma_i \cdot g_i$. Otherwise the algorithm outputs $\sign{\psi_i}$. Note that if $\sign{\psi_i}$ is not $\eps$-close to $f$ then there exists $g \in G_{\psi_i}$ such that $|\la f - \psi_i, g \ra_D| \geq 4 \tau$ and, in particular, $|v(g) - \la \psi_i, g \ra_D | \geq 3\tau$.

We set $\psi_{i+1}$ to be the projection of $\psi'_{i+1}$ onto $\funcs$ or
$\psi_{i+1}(x) \triangleq P_1(\psi'_{i+1}(x))$ and then continue to the next iteration using $\psi_{i+1}$.


As we can see $\sign{\psi_i}$ is only output when $\sign{\psi_i}$ is $\eps$-close to $f$. Therefore in order to prove the desired bound on the number of queries it is sufficient to show that the algorithm will output $\sign{\psi_i}$ after an appropriate number of iterations. This is established via the following claim.
\begin{claim}
\label{claim:count-steps}
For every $i$, $\|f - \psi_i \|_D^2 \leq 1 - 3 \cdot i \cdot \tau^2$.
\end{claim}
\begin{proof}
First, $\|f -\psi_0 \|_D^2 = \|f \|_D^2 = 1$. Next,
$$\|f -\psi'_{i+1} \|_D^2 = \| (f -\psi_{i}) - \gamma_i \cdot g_i \|_D^2 = \|f -\psi_{i} \|_D^2 +  \|\gamma_i \cdot g_i \|_D^2 - 2 \la f -\psi_{i}, \gamma_i \cdot g_i \ra_D .$$ Therefore,
\alequn{\|f -\psi_{i} \|_D^2 - \|f -\psi'_{i+1} \|_D^2 &= 2\gamma_i \la f-\psi_i, g_i \ra_D  - \gamma_i^2 \|g_i\|_D^2 \geq 2 \cdot  \gamma_i \cdot \la f - \psi_i, g_i \ra_D - \gamma_i^2 \\
& =^{(*)} 2 \cdot  |\gamma_i| \cdot |\la f - \psi_i, g_i \ra_D| - \gamma_i^2 \geq 2\cdot |\gamma_i| (|\gamma_i|-\tau) - \gamma_i^2 \geq \gamma_i^2/3 \geq 3 \cdot \tau^2 .}
To obtain $(*)$ we note that $|\gamma_i| \geq 3\tau$ and $|\la f - \psi_i, g_i \ra_D - \gamma_i| = |\la f, g_i \ra_D - v(g_i)|  \leq \tau$. Therefore the sign of $\gamma_i$ is the same as the sign of $\la f - \psi_i, g_i \ra_D$ and $|\la f - \psi_i, g_i \ra_D| \geq |\gamma_i| - \tau \geq 2\gamma_i/3$.

We now claim that $\|f -\psi'_{i+1} \|_D^2 \geq \|f -\psi_{i+1} \|_D^2$. This follows easily from the definition of
$\psi_{i+1}$. If for a point $x$, $\psi_{i+1}(x) = \psi'_{i+1}(x)$ then clearly $f(x)-\psi'_{i+1}(x) =  f(x)-\psi_{i+1}(x)$. Otherwise, if $|\psi'_{i+1}(x)| > 1$ then $\psi_{i+1}(x) = \sign{\psi'_{i+1}(x)}$ and for any value $f(x) \in \pmi$, $|f(x)-\psi'_{i+1}(x)| \geq  |f(x)-\psi_{i+1}(x)|$. This implies that $\E_D[(f -\psi'_{i+1})^2] \geq \E_D[(f -\psi_{i+1})^2]$.

We therefore obtain that for every $i$, $\|f -\psi_{i} \|_D^2 - \|f -\psi_{i+1} \|_D^2 \geq 3\tau^2$ giving us the claim.
\end{proof} (Cl.~\ref{claim:count-steps})

Claim \ref{claim:count-steps} implies that the algorithm makes at most $1/(3\tau^2)$ iterations. In each iteration at most $d$ queries are made and therefore the algorithm uses at most $d/(3\tau^2)$ queries of tolerance $\tau$.
\end{proof} (Th.~\ref{th:ssq2alg})

An important property of the proofs of Theorems \ref{th:learn2SQSD} and \ref{th:ssq2alg} that they give a simple and {\em efficient} way to convert a learning algorithm for $C$ into an algorithm that given access to target-independent statistical queries with respect to $D$ builds an approximating set $G_\psi$ for every $\psi$ and vice versa. As it was noted in \cite{Feldman:08ev}, the access to target-independent statistical queries with respect to $D$ can be replaced by a circuit that provides random samples from $D$ if $D$ is efficiently samplable or a fixed polynomial-size random (unlabeled) sample from $D$. In this case the resulting algorithm is non-uniform because it requires the random sample to be given to it as advice (see \cite{Feldman:08ev} for more details on converting a SQ algorithm to a CSQ algorithm). For convenience we refer to either of these options as access to $D$.
\begin{theorem}
\label{th:sqd-sqlearn-eff}
Let $C$ be a concept class and $D$ be a distribution over $X$. $C$ is {\em efficiently} SQ learnable over $D$ if and only if there exists an algorithm $\B$ that for every $\eps >0$ and $\psi \in \funcs$, given $\eps$, access to $D$ and a circuit for $\psi \in \funcs$ can produce a set of functions $G_\psi$ such that
\begin{enumerate}
\item $G_\psi$ satisfies the conditions of Definition \ref{def:ssqd} for some polynomial $d$ and inverse-polynomial $\gamma$ (in $n$, $1/\eps$);
\item circuit size of every function in $G_\psi$ is polynomial in $n$ and $1/\eps$;
\item the running time of $\B$ is polynomial in $n$, $1/\eps$ and the circuit size of $\psi$.
\end{enumerate}
\end{theorem}
\begin{proof}
The proof of Theorem \ref{th:learn2SQSD} gives a way to construct the set $G_\psi$ by simulating $\A$ while using $\psi$ in place of the target function $f$. This construction of $G_\psi$ would be efficient provided the exact values of $\E_D[\phi_i(x)\cdot \psi(x)]$ and the exact values of target-independent SQs in the simulation of algorithm $\A$ were available. However it is easy to see that the exact values are not necessary and can be replaced by estimates within $\tau/2$. Such estimates can be easily obtained given access to $D$.

Similarly, in the proof of Theorem \ref{th:ssq2alg} the iterative procedure would yield an efficient SQ learning algorithm for $C$ provided the exact values of $\la \psi_i, g \ra_D$ were available. In place of the exact values estimates within $\tau/2$ can be used if the accuracy of statistical queries is also increased to $\tau/2$. This implies that if there exists an efficient algorithm that given a polynomial size circuit for $\psi \in \funcs$ and access to $D$ generates $G_\psi$ then $C$ is efficiently SQ learnable over $D$.
\end{proof}

\subsection{Orthogonality-Based Characterization}
 In order to simplify the application of our characterization we show that, with only a polynomial loss in the bounds one can obtain an orthogonality-based version of \SQSD. Specifically, we convert the bound on the number of functions required to weakly approximate every function in some set of functions $F$ to a bound on the maximum number of almost uncorrelated functions in $F$.

First we extend the definition of SQ-DIM (Def.~\ref{def:old-weak-sqd}) to sets of arbitrary real-valued functions.
\begin{definition}
\label{def:extend-sqd}
For a set of real-valued functions $F$ we say that $\SQDIM(F,D) = d$ if $d$ is the largest value for which there exist $d$ functions  $f_1, f_2, \ldots, f_d \in F$ such that for every $i\neq j$,  $|\la f_i , f_j \ra_D| \leq 1/d.\ $
\end{definition}
Now state Yang's conversion (Lemma \ref{lem:SQD2SSQDIM}) generalized to sets of bounded real-valued functions. While it was stated in \cite{Yang:05} only for Boolean functions the only property of Boolean functions used in his proof is their $\|\cdot\|_D$-norm being equal to 1 (the same is also true and easier to verify in the simple proof by Sz\"{o}r\'{e}nyi \cite{Szorenyi:09}).
\begin{lemma}
Let $D$ be a distribution and $F$ be set of functions such that every $\phi \in F$, $\|\phi\|_D \leq 1$. Then
 $\SQD(F,D, d^{-1/3}) \geq d^{1/3}/2$,
where $d=\SQDIM(F,D)$.
\label{lem:approx2orthogonal}
\end{lemma}

We define $\SQSDIM(C,D,\eps)$ to be the generalization of \SQDIM\ to $\eps$-accurate learning as follows.
\begin{definition}
\label{def:strong-sqd}
$\SQSDIM(C,D,\eps) = \sup_{\psi \in \funcs} \SQDIM(C\setminus B^D(\sign{\psi},\eps) -\psi),D)$.
\end{definition}

We now ready to relate \SQSD\ and \SQSDIM.
\begin{theorem}
\label{th:ssqdim2sqd}
Let $C$ be a concept class $D$ be a distribution over $X$, $\eps > 0$ and $d = \SQSDIM(C,D,\eps)$. Then
$\SQSD(C,D,\eps,1/(2d)) \leq d$ and $\SQSD(C,D,\eps,d^{-1/3}) \geq d^{1/3}/4$.
\end{theorem}
\begin{proof}
Let $\psi \in \funcs$ be any function, let $F_\psi = C\setminus B^D(\sign{\psi},\eps) - \psi$ and let $d' = \SQDIM(F_\psi,D) \leq \SQSDIM(C,D,\eps) = d$.

For the first part of the claim we use a minor modification of the standard relation between \SQD\ and \SQSDIM (see Section \ref{sec:weaksqd}). Let $F_1 = \{f_1, f_2, \ldots, f_{d'}\} \subseteq F_\psi$ be a largest-size set of functions such that for every $i\neq j$, $|\la f_i , f_j \ra_D| \leq 1/d'.\ $ The maximality of $d'$ implies that for every $f \in F_\psi$, there exists $f_i \in F_1$ such that $|\la f_i , f \ra_D| > 1/d'.\ $ Thus $F_1$ is an approximating set for $F_\psi$. The only minor problem is that we need an approximating set of functions in $\funcs$. The domain of each function in $F_\psi$ is $[-2,2]$ and therefore to obtain an approximating set in $\funcs$ we simply scale $F_1$ by $1/2$. By taking $G_\psi = \{ f/2  \cond f \in F_1\}$ we obtain that $\SQD(F_\psi,D, 1/(2d')) \leq d'$. This holds for every $\psi \in \funcs$ and therefore $\SQSD(C,D,\eps,1/(2d)) \leq d$.

For the second part of the claim we first observe that for every $f \in F_\psi$, $f = c - \psi$ for $c \in C$ and hence $\|f\|_D \leq 2$. Let $F'_\psi$ denote with each function scaled by $1/2$ factor (to ensure that the norms are upper-bounded by $1$). By Lemma \ref{lem:approx2orthogonal} we obtain $\SQD(F'_\psi,D,(d'/4)^{-1/3}) \geq (d'/4)^{1/3}/2$. This implies that $\SQD(F_\psi,D,2(d'/4)^{-1/3}) \geq (d'/4)^{1/3}/2$ and hence $\SQD(F_\psi,D,(d')^{-1/3}) \geq (d')^{1/3}/4$ and $\SQSD(C,D,\eps,d^{-1/3}) \geq d^{1/3}/4$.
\end{proof}

We can combine Theorem \ref{th:ssqdim2sqd} with the approximation-based characterization (Th.~\ref{th:learn2SQSD} and \ref{th:ssq2alg}) to obtain a characterization of strong SQ learnability based on \SQSDIM.
\begin{theorem}
\label{th:ssqd-characterize}
Let $C$ be a concept class, $D$ be a distribution over $X$ and $\eps > 0$. If there exists a polynomial $p(\cdot,\cdot)$ such that $C$ is SQ learnable over $D$ to accuracy $\eps$ from $p(n,1/\eps)$ queries of tolerance $1/p(n,1/\eps)$ then $\SQSDIM(C,D,\eps + 1/p(n,1/\eps)) \leq p'(n,1/\eps)$ for some polynomial $p'(\cdot,\cdot)$. Further, if $\SQSDIM(C,D,\eps) \leq p(n,1/\eps)$ then $C$ is SQ learnable over $D$ to accuracy $\eps$ from $p'(n,1/\eps)$ queries of tolerance $1/p'(n,1/\eps)$ for some polynomial $p'(\cdot,\cdot)$.
\end{theorem}

\section{SQ Dimension for Agnostic Learning}
In this section we extend the statistical query dimension characterization to agnostic learning. Our characterization is based on the well-known observation that agnostic learning of a concept class $C$ requires (a weak form of) learning of the set of functions $F$ in which every function is weakly approximated by some function in $C$ \cite{KearnsSS:94}. For example agnostic learning of Boolean conjunctions implies weak learning of DNF expressions. We formalize this by defining an $L_1^D$ $\eps$-ball around a real-valued function $\phi$ over $X$ as $B_1^D(\phi,\eps) = \{\psi \in \funcs \cond  L_1^D(\psi,\phi) \leq \eps\}$ and around a set of functions $C$ as $B_1^D(C,\eps) = \cup_{f\in C}B_1^D(\phi,\eps)$. In $(\alpha,\beta)$-agnostic learning of a function class $C$ over the marginal distribution $D$, the learning algorithm only needs to learn when the distribution over examples $A=(D,\phi)$ satisfies $\Delta(A,C) \leq \alpha$. In other words, for any $A=(D,\phi)$ such that there exists $c \in C$, for which $\Delta(A,c) = L_1^D(\phi,c)/2 \leq \alpha$. Therefore $(\alpha,\beta)$-agnostic learning with respect to distribution $D$ can be seen as learning of the set of distributions $\D = \{ (D,\phi) \cond \phi \in B_1^D(C,2\alpha) \}$ with error of at most $\beta$. This observation allows us to apply the characterizations from Section \ref{sec:strong-sqd} after the straightforward generalization of \SQSD\ and \SQSDIM\ to general sets of real-valued functions. Namely, for a set of real-valued functions $F$, we define $$\SQSD(F,D,\eps,\gamma) = \sup_{\psi \in \funcs}\left\{\SQD(F\setminus B_1^D(\sign{\psi},2 \eps)- \psi,D,\gamma) \right\}\ .$$ The $\SQSDIM(F,D,\eps)$ is defined analogously. It is easy to see that when $F$ contains only $\pmi$ functions these generalized definitions are identical to Definitions \ref{def:ssqd} and \ref{def:strong-sqd}.

We can now characterize the query complexity of $(\alpha,\beta)$-agnostic SQ learning using $\SQSD(B_1^D(C,2 \cdot \alpha),D,\beta,\gamma)$ in exactly the same way as SLC is characterized using $\SQSD(C,D,\eps,\gamma)$. Formally, we obtain the following theorem.
\begin{theorem}
\label{th:learn2ASQSD}
Let $C$ be a concept class, $D$ be a distribution $D$ over $X$ and $0 < \alpha \leq \beta \leq 1/2$. Let $d$ be the smallest number of SQs of tolerance $\tau$ sufficient to $(\alpha,\beta)$-agnostically learn $C$. Then
\begin{enumerate}
\item $d \geq \SQSD(B_1^D(C,2 \cdot \alpha),D,\beta+\tau,\tau) - 2$,
\item $d \leq \SQSD(B_1^D(C,2 \cdot \alpha),D,\beta,4 \cdot \tau)/(3\tau^2)$.
\end{enumerate}
\end{theorem}
To prove Theorem \ref{th:learn2ASQSD} we only need to observe that the proofs of Theorems \ref{th:learn2SQSD} and \ref{th:ssq2alg} do not assume that the concept class $C$ contains only Boolean functions and hold for any class of functions contained in $\funcs$. To obtain a characterization of $(\alpha,\beta)$-agnostic SQ learning using \SQSDIM\ extend Theorem \ref{th:ssqdim2sqd} to general sets of functions in $\funcs$ (the proof can be used verbatim for this settings).
\begin{theorem}
\label{th:assqdim2asqd}
Let $F \subseteq \funcs$ be a set of functions, $D$ be a distribution over $X$, $\eps > 0$ and $d = \SQSDIM(F,D,\eps)$. Then $\SQSD(F,D,\eps,1/(2d)) \leq d$ and $\SQSD(F,D,\eps,d^{-1/3}) \geq d^{1/3}/4$.
\end{theorem}

While we can now use \SQSD\ or \SQSDIM\ to characterize SQ learnability in the basic agnostic model\footnote{This is the approach we used in the earlier version of this work.} a simpler approach to characterization is suggested by recent distribution-specific agnostic boosting algorithms \cite{Feldman:10ab,KalaiKanade:09}. Formally, a {\em weak agnostic} learning algorithm is an algorithm that can recover at least a polynomial fraction of the advantage over the random guessing of the best approximating function in $C$. Specifically, on a distribution $A = (D,\phi)$ it produces a hypothesis $h$ such that $\la h, \phi \ra_D \geq p(1/n,1-2\Delta(A,C))$ for some polynomial $p(\cdot,\cdot)$. Distribution-specific agnostic boosting algorithms of Kalai and Kanade \cite{KalaiKanade:09} and Feldman \cite{Feldman:10ab} imply the equivalence of weak and strong distribution-specific agnostic learning.
\begin{theorem}[\cite{Feldman:10ab,KalaiKanade:09}]
 \label{th:agnboost}
 Let $C$ be a concept class and $D$ be a distribution over $X$. If $C$ is efficiently weakly agnostically learnable over $D$ then $C$ is agnostically learnable over $D$.\end{theorem}
 This result is proved only for the example-based agnostic learning but, as with other boosting algorithms, it can be easily translated to the SQ model (\cf \cite{AslamDecatur:98b}). Given Theorem \ref{th:agnboost}, we can use the known characterizations of weak learning together with our simple observation to characterize the (strong) agnostic SQ learning using either SQD or SQ-DIM.
\begin{theorem}
\label{th:sqd4weak-agn}
Let $C$ be a concept class and $D$ be a distribution over $X$. There exists a polynomial $p(\cdot,\cdot)$ such that $\ASLC(C,D,\eps,1/p(n,1/\eps)) \leq p(n,1/\eps)$ if and only if there exists a polynomial $p'(\cdot,\cdot)$ such that for every $1> \Gamma > 0$, $\SQD(B_1^D(C,1-\Gamma), D, 1/p'(n,1/\Gamma)) \leq p'(n,1/\Gamma)$.
\end{theorem}
\begin{proof}
The proof is essentially the same as the characterization of weak learning by Bshouty and Feldman \cite{BshoutyFeldman:02}. We review it briefly for completeness. Given $\Gamma >0$ and an agnostic learning algorithm $\A$ for $C$, we simulate $\A$ with $\eps = \Gamma/4$ as in the proof of Theorem \ref{th:learn2SQSD} for $\psi \equiv 0$. Let $G$ be the set containing the correlational queries obtained from $\A$ and the final hypothesis. By the same analysis as in the proof of Theorem \ref{th:learn2SQSD}, the size of $G$ is upper-bounded by a polynomial in $n$ and $1/\eps=4/\Gamma$. Further, for every $\phi \in B_1^D(C,1-\Gamma)$, there exists $g \in G$ such that $|\la g, \phi \ra_D| \geq \min\{\tau, \Gamma-2\eps\} = \min\{\tau, \Gamma/2\}$. The tolerance of the learning algorithm is lower bounded by the inverse of a polynomial (in $n$ and $1/\Gamma$) and therefore we obtain the first direction of the claim.

If for every $\Gamma > 0$, $\SQD(B_1^D(C,1-\Gamma), D, 1/p'(n,1/\Gamma)) \leq p'(n,1/\Gamma)$ then $C$ can be weakly agnostically SQ learned by the following algorithm. First, ask the query $g \cdot \ell$ with tolerance $1/(3p'(n,1/\Gamma)$ for each function $g$ in the approximating set $G$. Let $v(g)$ denote the answer to the query for $g$. For a distribution $A = (D,\phi)$, $\E_A[g(x) \cdot b] = \la g, \phi \ra_D$ and therefore $|v(g) - \la g, \phi \ra_D| \leq 1/(3p'(n,1/\Gamma)$. By choosing $g' = \mbox{argmax}_{g\in G}\{|v(g)|\}$ we are guaranteed that $|\la g', \phi \ra_D| \geq 1/(3p'(n,1/\Gamma))$. Therefore $\sign{v(g')}\cdot g'$ is a weak hypothesis for $f$. Finally, we can appeal to Theorem \ref{th:agnboost} to convert this weak agnostic learning algorithm to a strong agnostic learning algorithm for $C$ over $D$.
\end{proof}

As before, we can now obtain an SQ-DIM--based characterization from the SQD--based one.
\begin{theorem}
\label{th:agn-sqdim}
Let $C$ be a concept class and $D$ be a distribution over $X$. There exists a polynomial $p(\cdot,\cdot)$ such that $\ASLC(C,D,\eps,1/p(n,1/\eps)) \leq p(n,1/\eps)$ if and only if there exists a polynomial $p'(\cdot,\cdot)$ such that for every $1>\Gamma > 0$, $\SQDIM(B_1^D(C,1-\Gamma), D) \leq p'(n,1/\Gamma)$.
\end{theorem}
\begin{proof}
Let $d=\SQDIM(B_1^D(C,1-\Gamma), D)$. Lemma \ref{lem:approx2orthogonal} implies that $\SQD(B_1^D(C,1-\Gamma), D, d^{-1/3}) \geq d^{1/3}/4$. This implies that $d \leq p_1(\SQD(B_1^D(C,1-\Gamma), D, 1/p_2(n,1/\Gamma)),1/\Gamma)$ for some polynomials $p_1(\cdot,\cdot)$ and $p_2(\cdot,\cdot)$. As in the case of concept classes, it follows immediately from the definition that $d \geq \SQD(B_1^D(C,1-\Gamma), D, 1/d)$. These bounds together with Theorem \ref{th:sqd4weak-agn} imply the claim.
\end{proof}

We now give a simple example of the use of this characterization. For $X = \zon$, let $U$ denote the uniform distribution over $\zon$ and let $C_{n,k}$ denote the concept class of all monotone conjunctions of at most $k$ Boolean variables.
\begin{theorem}
\label{th:agn-conjunction-lb}
For every $k=\omega(1)$, the concept class $C_{n,k}$ is not efficiently agnostically SQ learnable over the uniform distribution $U$.
\end{theorem}
\begin{proof}
Let $\chi_T$ denote the parity function of the variables with indices in $T \subseteq [n]$. Let $c_T$ denote the monotone conjunction of the same set of variables. If $|T|$ is odd then $\pr_U[\chi_T(x) \neq c_T(x)] = 1/2 - 2^{-|T|}$ and therefore $L_1^U(\chi_T,c_T) = 1-2^{-|T|+1}$. Similarly, for even $|T|$, $L_1^U(-\chi_T,c_T) = 1-2^{-|T|+1}$.  In particular, for $P_{n,k} = \{ (-1)^{|T|+1} \cdot \chi_T \cond |T| \leq k \}$, we get $P_{n,k} \subseteq B_1^U(C_{n,k},1-2^{-k+1})$. For any two distinct parity functions $\chi_S$ and $\chi_T$, $\la \chi_S,\chi_T \ra_U = 0$ and therefore $\SQDIM(B_1^U(C_{n,k},1-2^{-k+1}), U) \geq |P_{n,k}| = \sum_{i \leq k}{n \choose i}$. By choosing $\Gamma = 1/n$ we obtain that $\SQDIM(B_1^U(C_{n,k},1-\Gamma), U) = n^{\omega(1)}$. Theorem \ref{th:agn-sqdim} now implies the claim.
\end{proof}

Our proof shows that agnostic SQ learning of monotone disjunctions is hard because it requires weak SQ learning of example distributions that represent parity functions over the uniform distribution. Parity functions over the uniform distribution are well-known to be not weakly SQ learnable \cite{BlumFJ+:94}. An analogous approach was used by Kalai \etal to show that agnostic learning of majorities over the uniform distribution requires learning of parities with random noise which is a notoriously hard open problem in theoretical computer science \cite{KalaiKMS:08}. Their result also implies hardness of agnostic SQ learning of majorities and our result can also be seen as a reduction to learning of noisy parities. As far as we are aware, these are the only hardness results for agnostic learning of simple classes over the uniform distribution. A brief survey of other hardness results for agnostic learning can be found in \cite{FGKP:09}.

\section{Applications to Evolvability}
\label{sec:apply-evolve}
In this section we use the characterization of SQ learnability and the analysis in the proof of Theorem \ref{th:ssq2alg} to derive a new type of evolution algorithms in Valiant's framework of evolvability \cite{Valiant:09}.
\subsection{ Overview of the Model}
We start by presenting a brief overview of the model. For a detailed description and intuition behind the various choices made in model the reader is referred to \cite{Valiant:09,Feldman:09robust}. The goal of the model is to specify how organisms can acquire complex mechanisms via a resource-efficient process based on random mutations and guided by performance-based selection. The mechanisms are described in terms of the multi argument functions they implement.  The performance of such a mechanism is measured by evaluating the agreement of the mechanism with some ``ideal" behavior function. The value of the ``ideal" function on some input describes the most beneficial behavior for the condition represented by the input. The evaluation of the agreement with the ``ideal" function is derived by evaluating the function on a moderate number of inputs drawn from a probability distribution over the conditions that arise. These evaluations correspond to the experiences of one or more organisms that embody the mechanism.

Random variation is modeled by the existence of an explicit algorithm that acts on some fixed representation of mechanisms and for each representation of a mechanism produces representations of mutated versions of the mechanism. The model requires that the mutation algorithm be efficiently implementable. Selection is modeled by an explicit rule that determines the probabilities with which each of the mutations of a mechanism will be chosen to ``survive" based on the performance of all the mutations of the mechanism and the probabilities with which each of the mutations is produced by the mutation algorithm.

As can be seen from the above description, a performance landscape (given by a specific ``ideal" function and a distribution over the domain), a mutation algorithm, and a selection rule jointly determine how each step of an evolutionary process is performed. A class of functions $C$ is considered evolvable if there exist a representation of mechanisms $R$ and a mutation algorithm $M$ such that for every ``ideal" function $f \in C$, a sequence of evolutionary steps starting from any representation in $R$ and performed according to the description above ``converges" in a polynomial number of steps to $f$. This process is essentially PAC learning of $C$ with the selection rule (rather than explicit examples) providing the only target-specific feedback. We now define the model formally using the notation from \cite{Feldman:09robust}.

\subsection{Definition of Evolvability}
The description of an evolution algorithm $\A$ consists of the definition of the representation class $R$ of possibly randomized hypotheses in $\funcs$ and the description of polynomial time mutation algorithm $M$ that for every $r \in R$ and $\eps > 0$ outputs a random mutation of $r$
\begin{definition}
\label{def:mut-alg}
A {\em evolution algorithm} $\A$ is defined by a pair $(R,M)$ where
\begin{itemize}
\item $R$ is a representation class of functions over $X$ with range in $[-1,1]$.
\item $M$ is a randomized algorithm that, given $r \in R$ and $\eps$ as input, outputs a representation $r_1 \in R$ with probability $\pr_\A(r,r_1)$. The set of representations that can be output by $M(r,\eps)$ is referred to as the {\em neighborhood} of $r$ for $\eps$ and denoted by $\Neigh_\A(r, \eps)$.
\end{itemize}
\end{definition}
A {\em loss function} $L$ on a set of values $Y$ is a non-negative mapping $L: Y \times Y \rightarrow \R^+$. $L(y,y')$ measures the ``distance" between the desired value $y$ and the predicted value $y'$. In the context of learning Boolean functions using hypotheses with values in $[-1,1]$ we only consider functions $L: \pmi \times [-1,1] \rightarrow \R^+$. Valiant's original model only considers Boolean hypotheses and hence only the disagreement loss (or Boolean loss) which is equal to $L_\Delta(y,y') = y \cdot y'$. It was shown in our earlier work \cite{Feldman:09robust} that such loss is equivalent to the {\em linear loss} $L_1(y,y') = |y'-y|$ over hypotheses with the range in $[-1,1]$. Here we use the {\em quadratic loss} $L_Q(y,y') = (y'-y)^2$ function. For a function $\phi \in \funcs$ its performance relative to loss function $L$, distribution $D$ over the domain and target function $f$ is defined\footnote{In general, for this definition to make sense the loss function has to satisfy several simple properties to which we refer as being {\em admissible} \cite{Feldman:09robust}. Both loss functions we consider here are admissible and therefore we omit an explicit discussion of the general assumptions.} as $$L\LPerf_f(\phi,D) = 1-2\cdot \E_D[L(f(x),\phi(x))]/L(-1,1)\ .$$ For an integer $s$, functions $\phi,f \in \funcs$ over  $X$, distribution $D$ over $X$ and loss function $L$, the {\em empirical performance} $L\LPerf_f(\phi,D,s)$ of $\phi$ is a random variable that equals $1- \fr{s} \frac{2}{L(-1,1)} \sum_{i \in [s]}L(f(z_i), \phi(z_i))$ for $z_1,z_2,\ldots,z_s \in X$ chosen randomly and independently according to $D$.

A number of natural ways of modeling selection were discussed in prior work \cite{Valiant:09,Feldman:09robust}. For concreteness here we use the selection rule used in Valiant's main definition in a slightly generalized version from \cite{Feldman:09robust}. In selection rule $\SelNB[L,t,p,s]$ $p$ candidate mutations are sampled using the mutation algorithm. Then beneficial and neutral mutations are defined on the basis of their empirical performance $L\LPerf$ in $s$ experiments (or examples) using tolerance $t$. If some beneficial mutations are available one is chosen randomly according to their relative frequencies in the candidate pool. If none is available then one of the neutral mutations is output randomly according to their relative frequencies. If neither neutral or beneficial mutations are available, $\perp$ is output to mean that no mutation ``survived".
\begin{definition}
\label{def:selnb}
For a loss function $L$, tolerance $t$, candidate pool size $p$, sample size $s$, selection rule $\SelNB[L,t,p,s]$ is an algorithm that for any function $f$, distribution $D$, evolution algorithm $\A=(R,M)$, a representation $r \in R$, accuracy $\eps$, outputs a random variable that takes a value $r_1$ determined as follows. First run $M(r,\eps)$ $p$ times and let $Z$ be the set of representations obtained. For $r' \in Z$, let $\pr_Z(r')$ be the relative frequency with which $r'$ was generated among the $p$ observed representations. For each $r' \in Z \cup \{r\}$, compute an empirical value of performance $v(r') = L\LPerf_f(r',D,s)$. Let $\Bene(Z) = \{r' \cond v(r') \geq v(r) + t\}$ and $\Neut(Z) = \{r' \cond |v(r') - v(r)| < t\}$. Then
\begin{itemize}
\item[(i)] if $\Bene(Z) \neq \emptyset$ then output $r_1 \in \Bene$ with probability $\pr_Z(r_1)/ \sum_{r' \in \Bene(Z)} \pr_Z(r')$;
\item[(ii)] if $\Bene(Z) = \emptyset$ and $\Neut(Z) \neq \emptyset$ then output $r_1 \in \Neut(Z)$ with probability $\pr_Z(r_1)/ \sum_{r' \in \Neut(Z)} \pr_Z(r')$.
\item[(iii)] If $\Neut(Z) \cup \Bene(Z) = \emptyset$ then output $\perp$.
\end{itemize}
\end{definition}

A concept class $C$ is said to be evolvable by an evolution algorithm $\A$ in a selection rule $\Sel$ over distribution $D$ if for every target concept $f \in C$, mutation steps as defined by $\A$ and guided by $\Sel$ will converge to $f$.

\begin{definition}
\label{def:evolvability-new}
For concept class $C$ over $X$, distribution $D$, evolution algorithm $\A$, loss function $L$ and a selection rule $\Sel$ based on $L\LPerf$ we say that the class $C$ is evolvable over $D$ by $\A$ in $\Sel$ if there exists a polynomial $g(n,1/\eps)$ such that for every $n$, $f \in C$, $\eps > 0$, and every $r_0 \in R$, with probability at least $1-\eps$, a sequence $r_0,r_1,r_2, \ldots$, where $r_i \leftarrow \Sel(f,D,\A,r_{i-1})$ will have $L\LPerf_f(r_{g(n,1/\eps)}, D) > 1-\eps$. \end{definition}
We say that  an evolution algorithm $\A$ evolves $C$ over $D$ in $\Sel$ {\em monotonically} if with probability at least $1-\eps$, for every $i \leq g(n,1/\eps)$, $L\LPerf_f(r_{i}, D) \geq L\LPerf_f(r_0, D)$, where $g(n,1/\eps)$ and $r_0,r_1,r_2, \ldots$ are defined as above. Note that since the evolution algorithm can be started in any representation, this is equivalent to requiring that with probability at least $1-\eps$, $L\LPerf_f(r_{i+1}, D) \geq L\LPerf_f(r_i, D)$ for every $i$.

As in PAC learning, we say that a concept class $C$ is evolvable in $\Sel$ if it is evolvable over all distributions by a single evolution algorithm (we emphasize this by saying {\em distribution-independently} evolvable). A more relaxed notion of evolvability requires convergence only when the evolution starts from a single fixed representation $r_0$. Such evolvability is referred to as evolvability {\em with initialization}.

\subsection{Monotone Distribution-Specific Evolvability from SQ Learning Algorithms}
\label{ssec:distrib-specific-evolve}
In our earlier work \cite{Feldman:09robust} it was shown that every SQ learnable concept class $C$ is evolvable in $\SelNB[L_Q,t,p,s]$ (that is the basic selection rule with quadratic loss) for some polynomials $p(n,1/\eps)$ and $s(n,1/\eps)$ and an inverse polynomial $t(n,1/\eps)$. The evolution algorithms obtained in that result do not require initialization but instead are based on a form of implicit initialization that involves gradual reduction of performance to 0 if the process of evolution is not started in some fixed $r_0$. Such ``deliberate" gradual reduction in performance is possible since (somewhat unnaturally) $\SelNB$ allows a reduction in performance of  up to $t$ in every step. Taking many such steps is used to reinitialize the evolution algorithm. Hence we consider the question of whether it is possible to evolve from any starting representation without the need for performance decreases, in other words, which concept classes are evolvable monotonically. In this section we show that for every fixed distribution $D$ and every concept class $C$ SQ learnable over $D$, there exists a quadratic-loss monotone evolution algorithm for $C$ over $D$.

The key element of the proof of this result is essentially an observation that the SQ algorithm that we designed in the proof Theorem \ref{th:ssq2alg} can be seen as repeatedly testing a small set of candidate hypotheses, and choosing one that reduces the $\|\cdot \|_D^2$ distance to the target function. Converting such an algorithm to an evolution algorithm is a rather straightforward process.
First we show that Theorem \ref{th:learn2SQSD} gives a way to compute a neighborhood of every function $\psi$ that always contains a function with performance higher than $\psi$ (unless the performance of $\psi$ is close to the optimum).
\begin{theorem}
\label{th:sqlearn2neighbor}
Let $C$ be a concept class over $X$ and $D$ be a distribution. If $C$ is {\em efficiently} SQ learnable over $D$ then there exists an algorithm $\N$ that for every $\eps >0$, given $\eps$, access to $D$ and a circuit for $\psi \in \funcs$ can produce a set of functions $N(\psi,\eps)$  such that
\begin{enumerate}
\item For every $f\in C$, there exists $\phi \in N(\psi,\eps)$ such that
$$ \|f -\phi \|_D^2 \leq \max \{\eps, \|f -\psi \|_D^2 - \theta(n,1/\eps)\} , $$
for an inverse-polynomial $\theta(\cdot,\cdot)$;
\item the size of $N(\psi,\eps)$ is polynomial in $n$ and $1/\eps$;
\item the circuit size of every function in $N(\psi,\eps)$ is (additively) larger than the circuit size of $\psi$ by at most a polynomial in $n$ and $1/\eps$;
\item the running time of $\N$ is polynomial in $n$, $1/\eps$ and the circuit size of $\psi$.
\end{enumerate}
\end{theorem}
\begin{proof}
We use Theorem \ref{th:sqd-sqlearn-eff} to obtain an algorithm $\B$ that given a circuit for $\psi$, accuracy parameter $\eps$ and access to $D$, efficiently constructs set $G_\psi$ of polynomial size for some inverse polynomial $\gamma(n,1/\eps)$. Let $G_\psi(\eps/4)$ be the output of $\B$ on $\psi$, $\eps/4$ and access to distribution $D$. Now let $$N(\psi,\eps) = \left\{P_1(\psi + \gamma \cdot g) \cond g \in G_\psi(\eps/4) \right\} \bigcup \left\{P_1(\psi - \gamma \cdot g) \cond g \in G_\psi(\eps/4) \right\} \cup \{\sign{\psi}\}\ .$$
By the properties of $G_\psi(\eps/4)$, for every $f \in C$, either there exists a function $g \in G_\psi(\eps/4)$ such that $| \la f- \psi, g \ra_D | \geq \gamma(n,4/\eps)$, or $\pr_D[f \neq \sign{\psi}] \leq \eps/4$. In the first case, by the analysis in the proof of Theorem \ref{th:ssq2alg}, $\psi_g = P_1(\psi + b \cdot \gamma(n,4/\eps) \cdot g)$ satisfies $\|f-\psi_g \|_D^2 \leq \|f-\psi \|_D^2 - \gamma(n,4/\eps)^2$ for $b =\sign{\la f-\psi, g \ra_D}$. In the second case, $\|\sign{\psi} - f \|_D^2 \leq 4 \cdot \eps/4 = \eps$. Theorem \ref{th:sqd-sqlearn-eff} also implies that the algorithm that we have defined satisfies the bounds in conditions (2)-(4).
\end{proof}

By definition, $L_Q\LPerf_f(r,D) = 1 - \|f-r\|^2_D/2$. Hence an immediate corollary of Theorem \ref{th:sqlearn2neighbor} is monotone evolvability of every SQ-learnable concept class in $\SelNB[L_Q,t,p,s]$ over any fixed distribution $D$.

\begin{theorem}
\label{th:evolve-LSQ}
Let $D$ be a distribution and $C$ be a concept class efficiently SQ learnable over $D$. There exist polynomials $p(n,1/\eps)$ and $s(n,1/\eps)$, an inverse polynomial $t(n,1/\eps)$ and an evolution algorithm $\A = (R, M)$ such that $C$ is evolvable monotonically by $\A$ over $D$ in $\SelNB[L_Q,t(n,1/\eps),p(n,1/\eps),s(n,1/\eps)]$. Here if $D$ is not efficiently samplable then $\A$ is a non-uniform algorithm.
\end{theorem}
\begin{proof}
Let $R$ be the representation class containing all circuits over $X$ and let $r$ be any representation in $R$. Given $r$ and $1/\eps$ the algorithm $M$ uses the algorithm $\N$ from Theorem \ref{th:sqlearn2neighbor} with parameters $r$ and $\eps$ to obtain $N(r,\eps)$. Let $\theta(n,1/\eps)$ denote the inverse-polynomial improvement guaranteed by Theorem \ref{th:sqlearn2neighbor}. The algorithm $\N$ requires access to distribution $D$ and can be simulated efficiently if $D$ is efficiently samplable or simulated using a fixed random sample of points from $D$ otherwise. In this case $\A$ might be a non-uniform algorithm (as we explained in Section \ref{sec:strong-sqd}). The algorithm $M$ outputs a randomly and uniformly chosen representation in $N(r,\eps)$. The efficiency of $\N$ implies that $M$ can be implemented efficiently.

In order for this evolution algorithm to work we need to make sure that a representation with the highest performance in $N(r,\eps)$ is present in the candidate pool and that the performance of each candidate mutation is estimated sufficiently accurately. We denote a representation with the highest performance by $r^*$. The bound on the number of generations that we are going to prove is $g(n,1/\eps) = 8/\theta(n,1/\eps)$. To ensure that $r^*$ is with probability at least $1-\eps/4$ in the candidate pool in every generation we set $p(n,1/\eps) = |N(r,\eps)| \cdot \ln{\frac{4 \cdot g(n,1/\eps)}{\eps}}$. To ensure that with probability at least $1-\eps/4$ in every generation the performance of each mutation is estimated within $\theta(n,1/\eps)/8$ we set $s(n,1/\eps) = c \cdot \theta(n,1/\eps)^{-2} \cdot  \log{\frac{8 \cdot p(n,1/\eps) \cdot g(n,1/\eps)}{\eps}}$ for a constant $c$ (obtained via the Hoeffding's bound). We set the tolerance of the selection rule to $t(n,1/\eps) = 3 \cdot \theta(n,1/\eps)/8$.

By the properties of $\N$, $$L_Q\LPerf_f(r^*,D) \geq \min\{L_Q\LPerf_f(r,D) + \theta(n,1/\eps)/2, 1-\eps/2\}.\ $$ If $L_Q\LPerf_f(r,D) \leq 1 - \eps$ then $L_Q\LPerf_f(r^*,D) \geq L_Q\LPerf_f(r,D) + \theta(n,1/\eps)/2$ (without loss of generality $\theta(n,1/\eps) \leq \eps$). In this case if $r^*$ is in the pool of candidates $Z$ and the empirical performance of every mutation in $Z$ is within $\theta(n,1/\eps)/8$ of the true performance then $\Bene_Z(r)$ is non-empty and for every $r' \in \Bene_Z(r)$, $L_Q\LPerf_f(r',D) \geq L_Q\LPerf_f(r,D) + \theta(n,1/\eps)/4$. In particular, the output of $\SelNB[L_Q,t(n,1/\eps),p(n,1/\eps),s(n,1/\eps)]$ will have performance at least $L_Q\LPerf_f(r,D) + \theta(n,1/\eps)/4$. The lowest initial performance is $-1$ and therefore, with probability at least $1-\eps/2$, after at most $g(n,1/\eps) = 8/\theta(n,1/\eps)$ steps a representation with performance at least $1-\eps$ will be reached.

We also need to establish that once the performance of at least $1-\eps$ is reached it does not decrease within $g(n,1/\eps)$ steps and also prove that the evolution algorithm is monotone.
To ensure this we modify slightly the mutation algorithm $M$. The algorithm $M'$ outputs a randomly and uniformly chosen representation in $N(r,\eps)$ with probability $\Delta = \eps/(2 \cdot g(n,1/\eps))$ and outputs $r$ with probability $1-\Delta$. We also increase $p(n,1/\eps)$ accordingly to ensure that $r^*$ is still in the pool of candidates with sufficiently high probability. This change does not influence the analysis when $\Bene_Z(r)$ is non-empty. If $\Bene_Z(r)$ is empty then, by the definition of $M'$, $\SelNB[L_Q,t(n,1/\eps),p(n,1/\eps),s(n,1/\eps)]$ will output $r$ with probability at least $1-\Delta$. That is in every step, either the performance improves or it does not change with probability at least $1-\Delta$. In particular, with probability at least $1-\eps/2$ the performance will not decrease during any of the first $g(n,1/\eps)$ generations.
\end{proof}

\subsection{Distribution-Independent Evolvability of Disjunctions}
A substantial limitation of the general transformation given in the previous section is that the evolution algorithm given there requires access to $D$ and hence only implies evolvability for a fixed distribution. In this section we show that for the concept class of disjunctions (and conjunctions) the ideas of the transformation in Section \ref{ssec:distrib-specific-evolve} can be used to derive a simple algorithm for distribution-independent monotone evolvability of disjunctions. An even simpler and more general algorithm based on these ideas is also given in our subsequent work \cite{Feldman:11ltf}.

As usual in distribution-independent learning, we can assume that the disjunction is monotone \cite{KearnsLV:94}. We represent a monotone disjunction by a subset $T \subset [n]$ containing the indices of the variables in the disjunction and refer to it as $t_T$. For every $i\in [n]$, let $x_i$ refer to the function that returns the value of the $i$-th coordinate of a point in $\zon$.

Given a current representation computing function $\phi \in \funcs$ we try to modify it in two ways. The first one is to add $\gamma \cdot x_i$ and project using $P_1$ for some $i \in [n]$ and $\gamma > 0$. The other one is to subtract $\gamma$ and project using $P_1$. The purpose of the first type of modification is to increase performance on points where the target disjunction equals to 1. It is easy to see that such steps can make the performance on such points as close to 1 as desired. The problem with such steps is that they might also add $\gamma \cdot x_i$ such that $x_i$ is not in the target disjunction and thereby decrease the performance on points where the target equals $-1$. We fix this by using the second type of modification. This modification increases the performance on points where the target equals $-1$ but may decrease the performance on points where the target equals 1. The reason why this combination of modifications will converge to a good hypothesis is that for the quadratic loss function the change in loss due to an update is larger on points where the loss is larger. Namely, $L_Q(y,y'+\Delta) = L_Q(y,y') - 2 \cdot \Delta \cdot (y-y') + \Delta^2$. This means that if the first type of modification can no longer improve performance then the second type will. We formalize this argument in the lemma below.
\begin{lemma}
\label{lem:improve-disj}
For $\phi \in \funcs$, let $N_\gamma(\phi) = \{P_1(\phi + \gamma \cdot x_i) \cond i\in [n]\} \cup \{\phi,P_1(\phi - \gamma)\}$.
There exist inverse polynomial $\tau(\cdot,\cdot)$ and $\gamma(\cdot,\cdot)$ such that for every distribution $D$ over $\zon$, every target monotone disjunction $f$, every $\eps > 0$ and every $\phi(x) \in \funcs$ there exists $\phi' \in N_{\gamma(n,1/\eps)}(\phi)$ for which $$L_Q\LPerf_f(\phi',D) \geq \min\{L_Q\LPerf_f(\phi,D) + \tau(n,1/\eps), 1-\eps\}\ .$$
\end{lemma}
\begin{proof}
Let $f = t_T$ denote the target monotone disjunction. By the definition $\|f-\phi\|_D^2 = 2 (1 - L_Q\LPerf_f(\phi,D))$. We denote the loss of $\phi$ when $f$ restricted to $1$ and $-1$ by $\Delta_1= \E_D[(f-\phi)^2 \cdot (f+1)/2 ]$ and $\Delta_{-1}= \E_D[(f-\phi)^2 \cdot (1-f)/2 ]$ respectively. Let $\gamma = \eps^{3/2}/21$ and $\tau = \gamma^4/(8n)$.
We split the analysis into several cases.
\begin{enumerate}
\item If $L_Q\LPerf_f(\phi,D) \geq 1-\eps$ then $\phi' = \phi$ satisfies the condition.
\item $L_Q\LPerf_f(\phi,D) \leq 1-\eps$ and $\Delta_1 \geq 2 \gamma^2$. In this case,
$$\Delta_1 \leq \pr_D[f(x)=1,\ \phi(x)\geq 1-\gamma] \cdot \gamma^2 + \pr_D[f(x)=1,\ \phi(x)< 1-\gamma] \cdot 4\ .$$
Therefore $$\pr_D[f(x)=1,\ \phi(x)< 1-\gamma] \geq (\Delta_1 - \gamma^2)/4 \geq \gamma^2/4\ .$$
The target function is a disjunction of at most $n$ variables therefore there exists $i \in T$ such that $\pr_D[x_i=1,\ \phi(x)< 1-\gamma] \geq \gamma^2/(4n)$. For such $i$, let $\phi' = P_1(\phi + \gamma \cdot x_i)$. Note that for every point $x$, the loss of $\phi'(x)$ is at most the loss of $\phi(x)$ while for every point where $x_i=1$ and $\phi(x)< 1-\gamma$ the loss of $\phi'(x)$ is smaller than the loss of $\phi(x)$ by at least $\gamma^2$. Therefore,
 $$\|f-\phi'\|_D^2 \leq \|f-\phi\|_D^2 - \gamma^2 \cdot \pr_D[x_i=1,\ \phi(x)< 1-\gamma]
  \leq \|f-\phi\|_D^2 - \frac{\gamma^4}{(4n)}\ .$$ This implies that $$L_Q\LPerf_f(\phi',D) \geq L_Q\LPerf_f(\phi,D) + \tau(n,1/\eps)$$ for $\tau$ defined as above.
\item $L_Q\LPerf_f(\phi,D) \leq 1-\eps$ and $\Delta_1 < 2 \gamma^2$. In this case $\Delta_{-1} \geq 2\eps - \Delta_1 > 3 \cdot \eps/2$. Let $\phi' = P_1(\phi -\gamma)$. We now upper bound the increase in error on points where $f=1$ and lower bound the decrease in error on points where $f=-1$. For the upper bound we have
    $$\E_D[(f-(\phi-\gamma))^2] \leq 2 \cdot \E_D[(f-\phi)^2] + 2 \cdot \gamma^2 ,$$ and therefore the increase in error when $f =1$ is at most $\Delta_1 + 2 \cdot \gamma^2 \leq 4 \cdot \gamma^2$. For the lower bound similarly to the previous case we get the inequality
$$\Delta_{-1} \leq \pr_D[f(x)=-1,\ \phi(x)\leq -1+\sqrt{\eps}/2] \cdot \eps/4 + \pr_D[f(x)=-1,\ \phi(x)> -1+\sqrt{\eps}/2] \cdot 4\ .$$
Therefore \equ{\pr_D[f(x)=-1,\ \phi(x)> -1+\sqrt{\eps}/2] \geq (\Delta_{-1} - \eps/4)/4 \geq \eps/4\ .\label{eq:min-prob}} On every point $x$ where $f(x)=-1$ and $\phi(x)> -1+\sqrt{\eps}/2$,  $$|f(x)-\phi'(x)|^2 \leq |f(x)-\phi(x)|^2 - (2 \gamma (\phi(x)-f(x)) - \gamma^2)
  \leq |f(x)-\phi(x)|^2 - 2\gamma \sqrt{\eps}/2 + \gamma^2\ .$$
By combining this with equation (\ref{eq:min-prob}) and our choice of $\gamma = \eps^{3/2}/21$ we get
$$\|f-\phi'\|_D^2 \leq \|f-\phi\|_D^2 - \frac{\eps}{4} \cdot (\gamma \sqrt{\eps} - \gamma^2) \leq \|f-\phi\|_D^2 - 5 \cdot \gamma^2\ .$$ Therefore in this case
$$L_Q\LPerf_f(\phi',D) \geq L_Q\LPerf_f(\phi,D) + (5 \cdot \gamma^2 - 4 \cdot \gamma^2)/2 \geq L_Q\LPerf_f(\phi,D) + \tau(n,1/\eps)\ .$$
\end{enumerate}
\end{proof}

The neighborhood $N_\gamma(\phi)$ can be computed efficiently and therefore Lemma \ref{lem:improve-disj} can be converted to an evolution algorithm in exactly the same way as it was done in Theorem \ref{th:evolve-LSQ}. This implies monotone and distribution-independent evolvability of disjunctions in $\SelNB[L_Q,t,p,s]$.
\begin{theorem}
\label{th:evolve-disj}
There exist polynomials $p(n,1/\eps)$ and $s(n,1/\eps)$, an inverse polynomial $t(n,1/\eps)$ and an evolution algorithm $\A = (R, M)$ such that for every distribution $D$ disjunctions are evolvable monotonically by $\A$ over $D$ in $\SelNB[L_Q,t(n,1/\eps),p(n,1/\eps),s(n,1/\eps)]$.
\end{theorem}

\section{Discussion and Further Work}
One natural question not covered in this work is whether and how our characterization can be applied to understanding of the SQ complexity of learning specific concept classes for which the previously known characterizations are not sufficient. As we explained in the introduction, one such example is learning of monotone functions. This question is addressed in a recent work \cite{FeldmanLS:11colt},
where the first lower bounds for SQ learning of depth-3 monotone formulas over the uniform distribution are derived using \SQSDIM.
The main open problem in this direction is evaluating the \SQSDIM\ of monotone DNF over the uniform distribution.

As we have mentioned, another way to see our proof of Theorem \ref{th:ssq2alg} is as a boosting algorithm that instead of using a weak learning algorithm on different distributions uses a weak learning algorithm on different target functions (specifically on $f-\psi_i$ at iteration $i$). This perspective turned out to be useful for understanding of boosting in the agnostic learning framework. In particular, it has lead to the distribution-specific boosting algorithm given in Theorem \ref{th:agnboost} and to a new connection between agnostic and PAC learning.

We also believe that the insights into the structure of SQ learning given in this work will be useful in further exploration of Valiant's model of evolvability. For example, Theorem \ref{th:sqlearn2neighbor} can also be used to obtain distribution-specific evolvability of every SQ-learnable concept class with only very weak assumptions on the selection rule such as $(t,\gamma)$-distinguishing defined in \cite{Feldman:09robust} (we will elaborate this point elsewhere). In a subsequent work \cite{Feldman:11ltf} we use some of the ideas from this work to show that the important concept class of linear threshold functions with a non-negligible margin is evolvable monotonically and distribution-independently in a broad family of loss functions that includes the quadratic loss. This substantially generalizes our results for disjunctions and gives a simpler analysis. In addition we prove in \cite{Feldman:11ltf} that conjunctions are not evolvable distribution-independently with the Boolean loss. This suggests that other loss functions need to be considered to achieve distribution independence for even such simple concept classes, justifying our use of the quadratic loss. Perhaps, the most interesting question in this direction is whether results analogous to Theorem \ref{th:evolve-LSQ} can also be obtained for distribution-independent evolvability and extended to other interesting loss functions.

In another related subsequent work Kanade \etal study monotonicity with the Boolean loss \cite{KanadeVV:10}. They show that strict monotonicity (which is satisfied by the algorithms we give here) implies robustness of the evolution algorithm to gradual change in the target function. They also give two new monotone evolution algorithms for linear threshold functions (with different assumptions on the distribution over the domain). Finally, in a very recent work P.~Valiant extended the model of evolvability to real-valued target functions \cite{Valiantp:11eccc}. Along with a general transformation of optimization algorithms to his new model, he described a simple evolution algorithm for monotone and distribution independent evolving of linear functions using, again, the quadratic loss function.

\section*{Acknowledgements}
I thank Nader Bshouty, Hans Simon and Les Valiant for discussions and valuable comments on this work. I am also grateful to the anonymous reviewers of FOCS 2009 and JCSS for a number of insightful comments and useful corrections.

\pagebreak
\bibliographystyle{plain}


\end{document}